# High Curie temperature half metallic 2D $M_2Se_3$ ($M$ = Co, Ni, and Pd) monolayers with superior mechanical flexibility

Peng Lv,[†] Gang Tang,[†] Chao Yang,[†] Jianming Deng,[†] Yanyu Liu,[†] Xueyun Wang,[†] Xianqiao Wang[‡] and Jiawang Hong[†*]

[†]*School of Aerospace Engineering, Beijing Institute of Technology, Beijing 100081, China*

[‡]*College of Engineering, University of Georgia, Athens, GA 30602, USA*

.

## ABSTRACT:

Pursuing two-dimensional (2D) intrinsic ferromagnetism with high Curie temperature and great mechanical flexibility has attracted great interest in flexible spintronics. In the present work, we carried out a density functional theory (DFT) investigation on the 2D $M_2Se_3$ ($M$ = Co, Ni, and Pd) monolayers to understand their structural stabilities, electronic, magnetic and mechanical properties. Our results show that the $Co_2Se_3$ monolayer exhibits a fascinating half-metallic ferromagnetism with high Curie temperature (>700 K). In addition, due to their unique buckling hinge-like structure, $M_2Se_3$ monolayers possess the large out-of-plane negative Poisson's ratio (NPR) and superior mechanical flexibility evidenced by their unusual critical strain (~50-60%) two times greater than the well-known 2D materials. These findings imply that 2D $M_2Se_3$ family is the promising materials for the applications in the flexible and high-density spintronic nanodevices.

[*] Corresponding author. E-mail: hongjw@bit.edu.cn (J Hong)

Flexible spintronics[1-3] are the key requirements for smart data processing, nonvolatile information storage, and radio frequency communication in the flexible and wearable electronic systems.[4-9] Some flexible spintronic nanomaterials, such as giant magnetoresistance[1, 10] or tunneling magnetoresistance[11] structures have been fabricated on flexible substrates, suggesting an important procedure towards flexible electronics. Since graphene has been discovered experimentally by Novoselov *et al.* in 2004,[12] various two-dimensional (2D) materials have attracted extraordinary attentions due to their fascinating physical and chemical properties.[13-17] The high-temperature ferromagnetism[18] (FM) and great mechanical flexibility[5] would render 2D materials as the promising and outstanding candidates for the flexible spintronic nanodevices.

In 2D materials, the local magnetism could be induced by lattice defect engineering or strain engineering. Recent works have suggested that the intrinsic defect,[19] adatom adsorption,[20] and substituted doping[19, 21-24] can induce the magnetic moments in graphene by breaking the delocalized $\pi$ bonding network. 2D materials of $NbS_2$ and $NbSe_2$ structures show magnetization under an external strain.[25] Very recently, several 2D ferromagnet studies[26-29] have shown experimentally that the intrinsic FM ordering persists down to the ultrathin limit, such as $CrI_3$[26, 28], $VSe_2$,[27] and $Cr_2Ge_2Te_6$.[29] All these works pave the path for exploring 2D intrinsic FM materials. However, such systems have generally low Curie temperature ($T_C$) and it is challenging to be used in the applications.

For the application of flexible spintronic devices, magnetic materials also call for the excellent mechanical properties to bear the large strain but still keep the original functionalities. Unusually, 2D materials have better flexibility than the bulk materials, which can be utilized in the flexible, stretchable and foldable electronics.[5] For example, graphene can sustain a large tensile strain up to 19% (critical strain) before its structural failure,[30] exhibiting extreme mechanical stability in comparison with conventional materials used in flexible electronics.[5, 31] To the best of our knowledge, well-known 2D material with the largest ideal strain is $MoS_2$ (36% along zigzag direction)[32]. However, the most 2D flexible materials are intrinsic nonmagnetic, which limits their applications in the flexible spintronic systems.

Therefore, it is crucial to design new 2D intrinsic FM materials with both high Curie temperature and outstanding mechanical properties. Recently, the Se-deficient $Pd_2Se_3$ monolayer with a novel structure was successfully synthesized,[33] which exhibits anisotropic transport properties and pronounced optical absorption.[34] However, its electronic, magnetic, and mechanical properties remain unexplored. Inspired by its unique structure, we constructed two other possible magnetic 2D systems, *i.e.* $Co_2Se_3$ and $Ni_2Se_3$, together with $Pd_2Se_3$, to investigate their structural, electronic, magnetic, and mechanical properties by using first-principles method, aiming at the high-temperature magnetic 2D materials with superior mechanical flexibility. Interestingly, we identify a 2D half-metallic material ($Co_2Se_3$) in the family of $M_2Se_3$ ($M$ = Co, Ni, and Pd) with high Curie temperature (>700 K). Moreover, $M_2Se_3$ family shows large critical strain and giant out-of-plane negative Poisson's ratio (NPR) due to its unique hinge-like structure. These results suggest that 2D $M_2Se_3$ family is the promising materials for the applications in the new generation of flexible spintronic nanodevices.

***Structural Stabilities.*** First, we examine the atomic structure of 2D $M_2Se_3$ monolayers. **Figure 1** shows the lattice structure of $M_2Se_3$ monolayers and **Table 1** summarizes the optimized lattice constants, bond lengths, buckling height, and cohesive energy of three structure-similar 2D monolayers: $Co_2Se_3$, $Ni_2Se_3$, and $Pd_2Se_3$. It can be found that the lattice constants (*a* and *b*) increase monotonically as the metal atomic radius increases. The monolayer structure is comprised of five layers along the thickness direction where the middle metal-layer is sandwiched by the outside Se-layers. Along the *y* (*b*) direction, two nearest Se atoms at the outermost layer form Se-dimer with the covalent bond ($d_1$). In the Se-dimer, each Se atom bonds to two metal atoms, marked as $d_2$. Other Se atoms that are not within the Se-dimer bond to four metal atoms, marked as $d_3$. Each unit cell is composed by two hinge-like structure (blue rectangle), connected directly by metal atoms. The electron localization function (ELF)[35, 36] and charge density distribution (please see **Figure S1** in the supplemental materials) is employed to understand the bonding nature of the $M_2Se_3$ monolayers, in which the ELF and charge density are mainly concentrated around the Se atoms and the corresponding Se-dimers are formed.

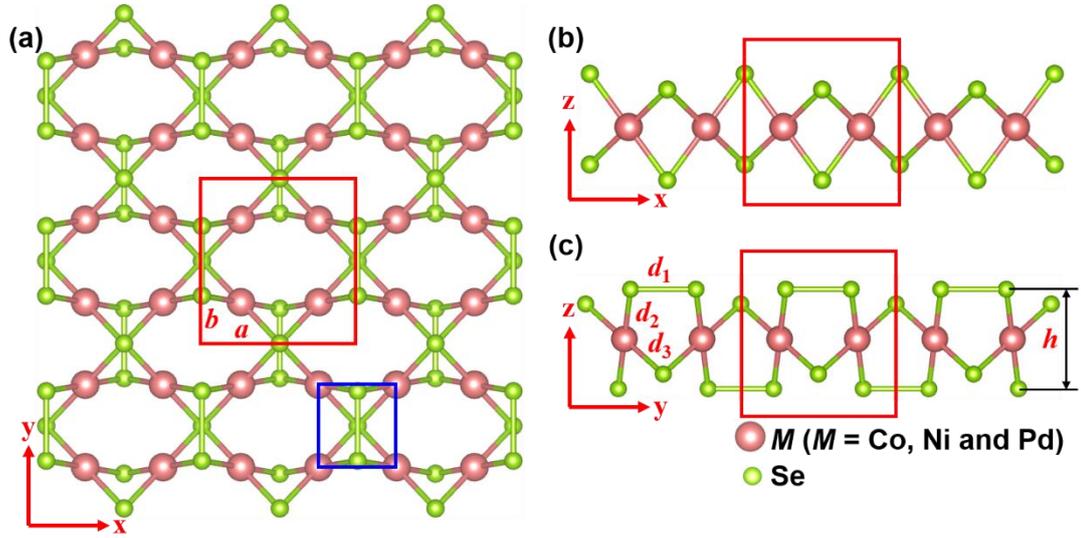

**Figure 1.** The atomic structures of 2D $M_2Se_3$ ($M$ = Co, Ni, and Pd) monolayers. (a) Top and (b, c) side views of the $M_2Se_3$ monolayers. The red rectangle with Bravais lattice vectors ($a$ and $b$) are the corresponding primitive unit cell. The blue rectangle shows the hinge-like structure.

**Table 1.** The structural parameters and energetics of 2D $M_2Se_3$ monolayers: lattice constants $a$ and $b$; bond lengths $d_1$, $d_2$, and $d_3$; buckling height $h$ and cohesive energy $E_c$.

| Structures | $a$ (Å) | $b$ (Å) | $d_1$ (Å) | $d_2$ (Å) | $d_3$ (Å) | $h$ (Å) | $E_c$ (eV/atom) |
|---|---|---|---|---|---|---|---|
| $Co_2Se_3$ | 5.415 | 5.782 | 2.436 | 2.320 | 2.379 | 3.740 | -3.84 |
| $Ni_2Se_3$ | 5.431 | 5.929 | 2.432 | 2.313 | 2.386 | 3.706 | -3.71 |
| $Pd_2Se_3$ | 5.984 | 6.135 | 2.403 | 2.461 | 2.544 | 3.852 | -3.32 |

In order to determine the ground state of the 2D $M_2Se_3$ monolayers, we calculated their total energies with different spin orders, as shown in **Figure S2**. Our calculations show that $Co_2Se_3$ monolayer has the FM state (see **Table S1**) while $Ni_2Se_3$ and $Pd_2Se_3$ monolayers have nonmagnetic (NM) state. Our results further show that the energy of the FM state of $Co_2Se_3$ is lower by at least 99.8 meV per unit cell than that of the antiferromagnetic (AFM) state.

The magnetic moment of Co is 1.04 $\mu_B$ per atom, according to the Bader-type charge analysis.[37]

Next, we focus on the structural stabilities of FM $Co_2Se_3$ and NM $Ni_2Se_3$, $Pd_2Se_3$ monolayers, according to their corresponding ground states. Previous experimental study showed that 2D $Pd_2Se_3$ monolayer could be successfully synthesized with great stability,[33] while there is no report about the fabrication of $Co_2Se_3$ and $Ni_2Se_3$ monolayers up to now. Therefore, it is useful to assess and compare structural stabilities of these 2D $M_2Se_3$ monolayers. The calculated cohesive energy $E_c$ for the $Co_2Se_3$, $Ni_2Se_3$ and $Pd_2Se_3$ are -3.84, -3.71 and -3.32 eV/atom (**Table 1**), respectively. The $E_c$ of $Co_2Se_3$ and $Ni_2Se_3$ monolayers are comparable to that of $Pd_2Se_3$ monolayer, which indicates the energetic stability and synthetic possibility of $Co_2Se_3$ and $Ni_2Se_3$ monolayers. The phonon dispersions of $M_2Se_3$ (**Figure S3**) show no negative phonon energy, which further confirms their dynamic stability. Furthermore, the AIMD simulations were performed to check the thermal stability of the 2D $M_2Se_3$ monolayers under ambient conditions. Note that the total free energy of $M_2Se_3$ monolayers only shows slightly fluctuation during the entire period of simulation (**Figure S4**). Simulation snapshots of the $M_2Se_3$ monolayers at 600 K also show their structural integrity up to 5 ps, implying their thermal stability.

***Electronic Properties.*** The band structures and partial density of states (PDOS) of 2D $M_2Se_3$ monolayers are calculated and present in **Figure 2**. Interestingly, it can be clearly seen that $Co_2Se_3$ shows the half-metal property with a moderate band gap (0.62 eV) for spin-up state (red lines) while the band can cross the Fermi level for spin-down state (blue lines) (**Figure 2b**). This half-metallicity results from Co atoms, as the spin-down channel beyond the Fermi level is mainly attributed to the Co-3$d$ states. **Figure 2a** depicts a wide range of Co-3$d$ states which accounts for the ferromagnetism.

For the NM $Ni_2Se_3$ and $Pd_2Se_3$ monolayers, they both show the indirect band gaps. The $Pd_2Se_3$ monolayer has an indirect band gap of 0.41 eV with the valence band maximum (VBM) lying at the Γ point and the conduction band minimum (CBM) locating along the S-Y direction (**Figure 2d**), which is in good agreement with the previous reports.[33, 34] The PDOS analysis reveals that the VBM of the Pd$_2$Se$_3$ monolayer mainly consists of Pd-4$d$ states, while the CBM is from both Pd-4$d$ and Se-4$p$ states with nearly equal weight. Similar to the Pd$_2$Se$_3$ monolayer, Ni$_2$Se$_3$ has an indirect band gap of 0.33 eV with its VBM locating along the Γ-X direction and CBM along the S-Y direction. The VBM is composed by Ni-3$d$ states, while the CBM is contributed by Se-4$p$ and higher Ni-3$d$ states.

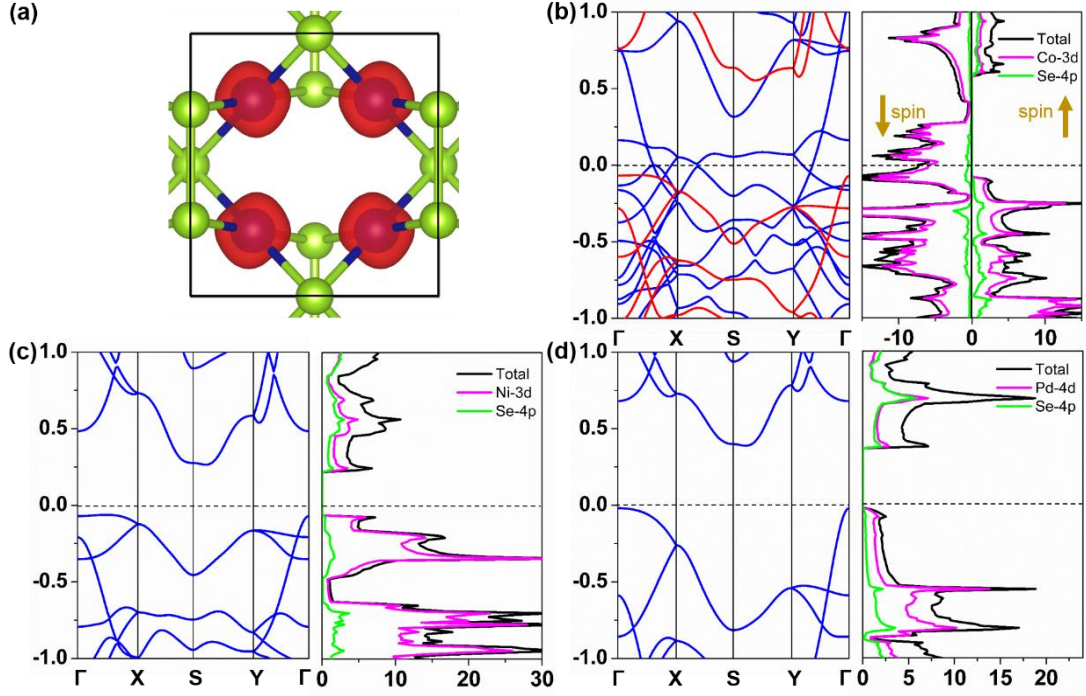

**Figure 2.** The spin density of Co$_2$Se$_3$ is shown in (a). The band structure and DOS are for (b) Co$_2$Se$_3$, (c) Ni$_2$Se$_3$ and (d) Pd$_2$Se$_3$ monolayers, respectively. The dashed line represents the Fermi level. The red and blue lines in the band structures for Co$_2$Se$_3$ (b) represent the spin-up and spin-down states, respectively.

*Magnetic Properties.* As discussed above, Co$_2$Se$_3$ monolayer is an interesting 2D half-metallic system with large intrinsic ferromagnetism. For the application of the half-metallic materials, it is desirable to have a ferromagnetic property with the Curie temperature $T_C$ higher than the room temperature. Therefore, it will be vital to check the magnetic stability of 2D Co$_2$Se$_3$ monolayer through the temperature effect. The $T_C$ was calculated from the mean field theory (MFT):[38, 39]

$$3k_B T_C = 2\Delta E \qquad (1)$$

where $k_B$ is the Boltzmann constant and $\Delta E$ is the energy difference between the FM and the lowest AFM state (AFM3). Taking the energy values from the **Table S1**, we estimate the $T_C$ of $Co_2Se_3$ monolayer as ~770 K. We also verified the $T_C$ of $Co_2Se_3$ monolayer based on the spin-polarized AIMD simulations at 600 K, 700 K and 800 K, as shown in **Figure S5**. It is clear that 2D $Co_2Se_3$ monolayer almost retains its original structure and magnetism up to 700 K, indicating both its thermal and magnetic stability above 700 K. As the temperature goes beyond 800 K, the structure loses its stability to fall apart accompanied with a significant decrease of the magnetic moment. Our results show that 2D $Co_2Se_3$ monolayer is a promising half-metallic material with robust ferromagnetism and structure up to 700 K.

We also investigated the magnetic stability of $Co_2Se_3$ monolayer under an external strain within the range of -5% to 5%. **Figure 3a** shows that the magnetic moment of $Co_2Se_3$ monolayer almost keeps constant under the compressive strain along the *x* direction or tensile strain along the *y* direction, which indicates that the half-metallic state is robust under such strain state. However, the magnetic moment decreases and the $Co_2Se_3$ monolayer turns into a FM metal when the strain is greater than +2% along the *x* direction (or less than -1% along the *y* direction). **Figure 3b** (**Figure 3c**) shows that applying +3% (-2%) strain along the *x* (*y*) direction leads to the VBM of spin-up state moving up through the Fermi level. These reveal that tensile strain (>2%) along the *x* direction or compressive strain (<-1%) along the *y* direction can eliminate the half-metallic state in 2D $Co_2Se_3$ monolayer. Therefore, these strain states should be avoided in order to keep the half-metallic FM state for the applications in the spintronic devices.

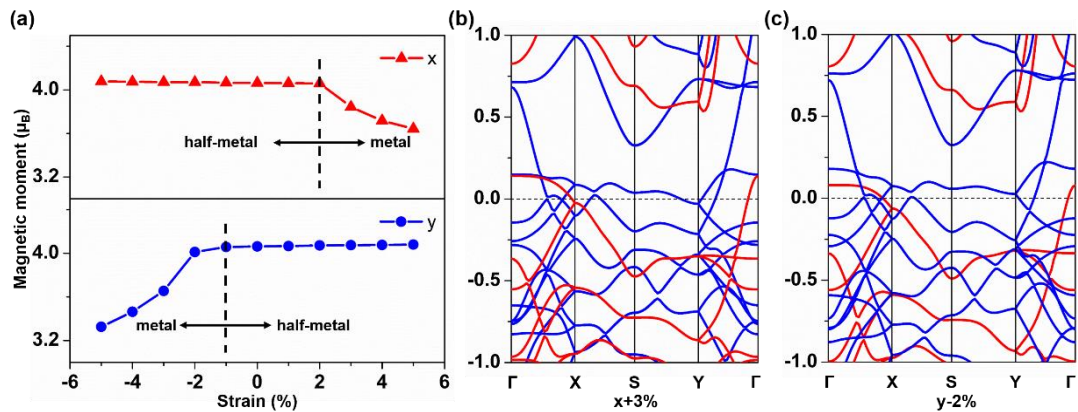

**Figure 3.** The magnetic moment (in $\mu_B$ per unit cell) of the $Co_2Se_3$ monolayer under (a) uniaxial strain. The band structure (b) under +3% strain along the *x* direction and (c) -2% strain along the *y* direction. The red and blue lines in the band structures represent the spin-up and spin-down states, respectively.

***Mechanical Properties***. We investigated the ideal tensile strength and critical strain of these $M_2Se_3$ monolayers based on the stress-strain relationship under uniaxial strain, as shown in **Figure 4**. The ideal strength and critical strain of $M_2Se_3$ along the *x* and *y* directions are summarized in **Table 2**. It can be seen that the ideal strength along the *x* direction are larger than that along the *y* direction for all three 2D monolayers. Moreover, the critical strain is found to be unusually large, which is about 2~4 times larger than that of many representative 2D materials (**Figure 5**). This large critical strain results from the unique buckling hinge-like structure of $M_2Se_3$, making their stretched deformation similar to the natural and biomimetic polymer with folded structures,[40-42] in which the unfolding of the coiled chains are mainly in response to mechanical deformation firstly, and then the bond effectively bears the tensile load before the structural failure. Under the external strain, 2D $M_2Se_3$ monolayers have the similar deformation process: bond angles change much faster than the bond lengths change ("unfold") in the first stage and then the bond lengths change dominates the next stage before the failure of monolayers. For example, under the tensile strain ( less than 20%) along the *y* direction (**Figures 6c** and **6d**), the change of the bond $d_3$ is 2% while the bond angle $\alpha$ opens by 22%. This implies that the dominating bond angles sustain the external strain. From 20% to the critical strain 53%, the bond angles $\alpha$ has the similar change (26%) to the first state, while the bond $d_3$ changes significantly fast by 12%, which is 6 times larger than that in the first state, indicating that the tensile strain extensively stretches the chemical bonds in the second stage before the failure of monolayers. The similar process can be found along the *x* direction (**Figures 6a** and **6b**) and also in $Ni_2Se_3$ and $Pd_2Se_3$ monolayers (**Figures S6** and **S7**). This "unfolding" mechanism resulting from the buckling hinge-like structure endows 2D $M_2Se_3$ monolayers with high convertibility and great flexibility, which provides an effective strategy to exploit 2D flexible nanomaterials.

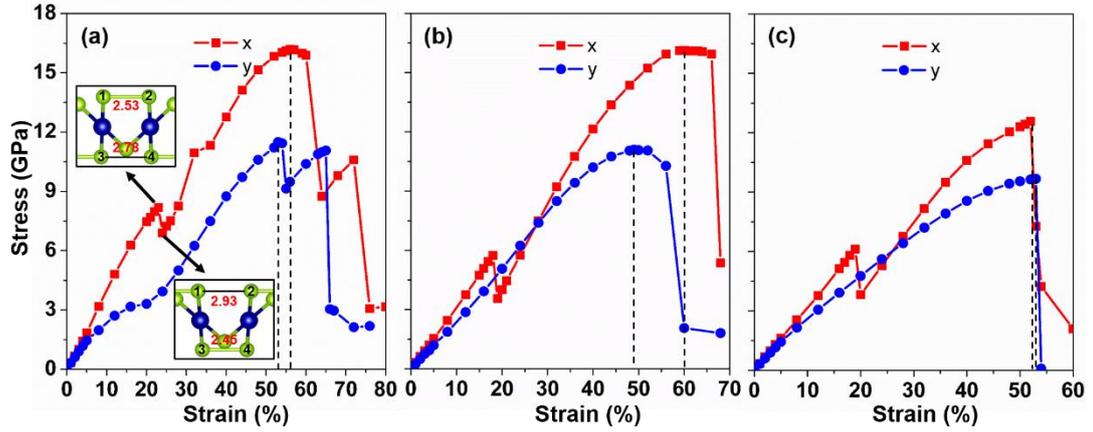

**Figure 4.** The stress-strain relationship along the $x$ and $y$ directions for (a) $Co_2Se_3$, (b) $Ni_2Se_3$, and (c) $Pd_2Se_3$ monolayers. The corresponding critical strains are labeled as the dashed line. The insets in (a) show the structural rearrangement when stretching along the $x$ direction at strain of 23% and 24%: Se1-Se2 bond breaks while Se3-Se4 bond forms. The red texts are the bond length in Å.

**Table 2.** The ideal strength and critical strain of $M_2Se_3$ along the $x$ and $y$ directions.

| Structures | Ideal strength (GPa) | | Critical strain | |
| --- | --- | --- | --- | --- |
| | $x$ | $y$ | $x$ | $y$ |
| $Co_2Se_3$ | 16.19 | 11.50 | 56% | 53% |
| $Ni_2Se_3$ | 16.12 | 11.10 | 60% | 49% |
| $Pd_2Se_3$ | 12.57 | 9.67 | 52% | 53% |

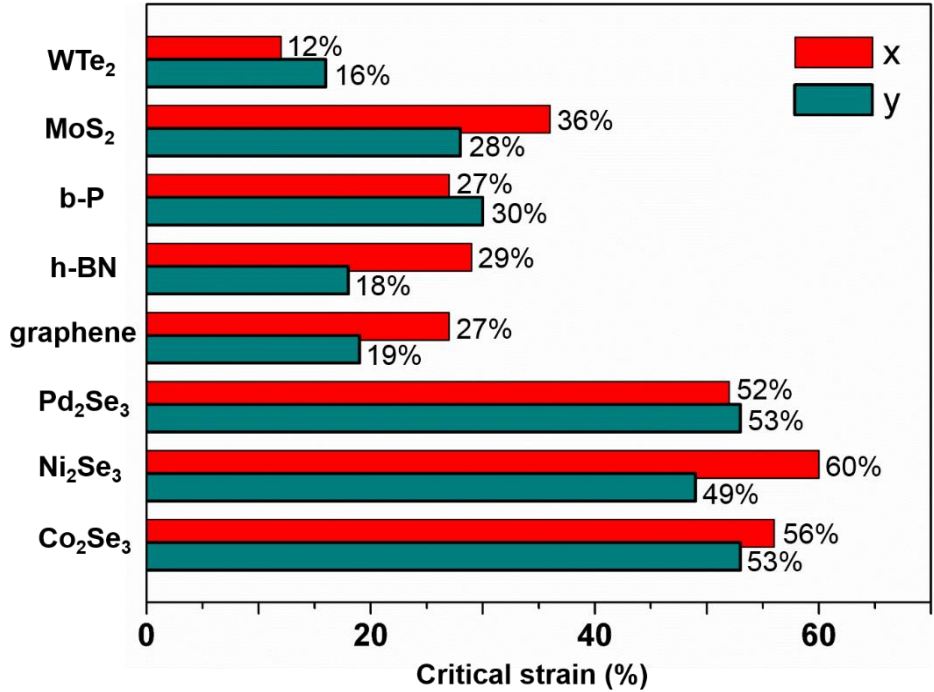

**Figure 5.** The critical strain along the $x$ and $y$ directions for $M_2Se_3$, along with the other typical 2D materials (graphene,[30] $h$-BN,[43] $b$-P,[44] $MoS_2$,[32] and $WTe_2$[45]) for the comparison.

Interestingly, it is worth noting that the stress suddenly jumps when stretching along the $x$ direction in $M_2Se_3$, and then continues to increase, as shown in **Figure 4**. To understand this phenomenon, we examine the crystal structure of $Co_2Se_3$ at different tensile strains along the $x$ direction, as shown in the insets in **Figure 4a**. From 0 to 23%, both Se1 and Se2 atoms in the Se-dimer (bond $d_1$) are moving away from each other while both Se3 and Se4 atoms (bond $d_4$) are drawing near to each other, as shown in **Figure 6a**. When tensile strain increases to 24%, the $d_1$ bond suddenly breaks while a new Se-dimer (Se3 and Se4) forms with a chemical bond ($d_4$) of 2.45 Å. The dimer breaking and reforming, termed as structural rearrangement, also occurs in $Ni_2Se_3$ and $Pd_2Se_3$ monolayers. This structural rearrangement resists the deformation and protects the structural integrity of hinge-like structures. When the strain is released, we found that the rearranged structure turns back to the original one, *i.e.*, $d_1$ bond forms again and $d_4$ bond breaks. This reversible structure under large strain is critical for the applications of flexible electronic devices.

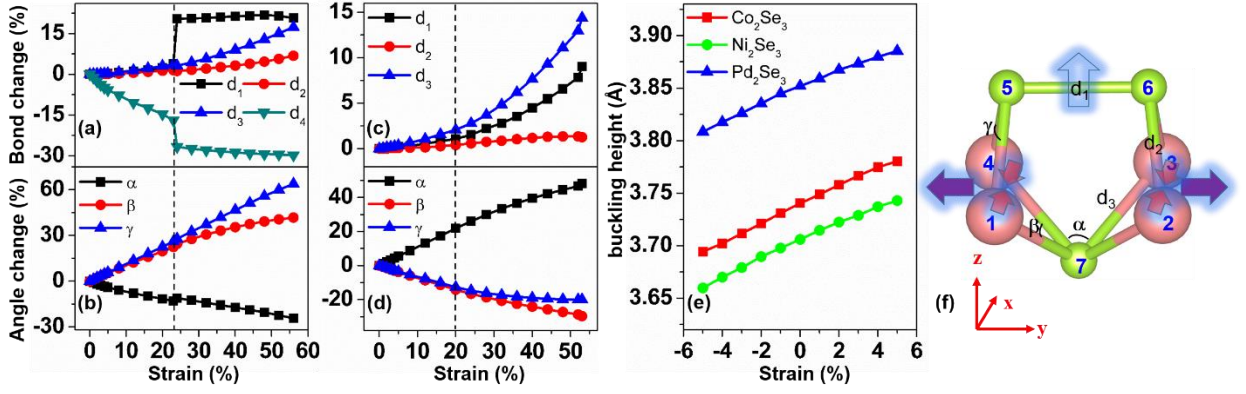

**Figure 6.** The change of the bond lengths and bond angles for the $Co_2Se_3$ monolayer under tensile strain along the $x$ (a, b) and $y$ (c, d) directions. The bond lengths and angles are labeled in (f). (e) The evolution of buckling height by a uniaxial strain along the $y$ direction for monolayer $M_2Se_3$, indicating the out-of-plane negative Poisson's ratio.

We calculated the in-plane Young's modulus of $M_2Se_3$ by linear fitting the stress versus strain with the strain less than 3% in **Figure 4**. The Poisson's ratios were also obtained after the structure relaxations with different strains (**Figure S8**). We also calculated the elastic constants (**Table S2**) of $M_2Se_3$ monolayers using the finite differences method.[46] Note that these values satisfy the requirements of the mechanical stability criterion of a 2D material:[47] $C_{11} \times C_{22} - C_{12}^2 > 0$ and $C_{66} > 0$, indicating that these $M_2Se_3$ monolayers are mechanically stable. The in-plane Young's modulus and Poisson's ratio were also calculated from the elastic constants (see Equations 3), which are summarized in **Table 3**. We can see that the mechanical properties from different methods agree well with each other. The results in **Table 3** show weak anisotropy of the mechanical properties, and the in-plane Young's modulus along the $x$ direction are slightly larger than that along the $y$ direction. It is noted that the Young's moduli of the $M_2Se_3$ are much smaller than that of other 2D materials, like graphene (~1000 GPa),[30, 48, 49] $h$-BN (250 GPa),[32, 50] $b$-P (166 GPa),[44] $MoS_2$ (330 GPa),[51] and $WTe_2$ (140 GPa),[45] suggesting that $M_2Se_3$ monolayers are indeed very soft and suitable for the flexible electronics.

**Table 3.** Mechanical properties of 2D $M_2Se_3$ monolayer. $E_x$ and $E_y$ (GPa) are the in-plane

Young's modulus, $v_{xy}$ and $v_{yx}$ ($v_{xz}$ and $v_{yz}$) are the in-plane (out-of-plane) Poisson's ratio. The values without (with) parentheses are calculated from stress-strain relationship (the finite differences method).

| Structures | $E_x$ | $E_y$ | $v_{xy}$ | $v_{xz}$ | $v_{yx}$ | $v_{yz}$ |
|---|---|---|---|---|---|---|
| $Co_2Se_3$ | 32.77 | 28.00 | 0.73 | 0.44 | 0.61 | -0.24 |
| | (33.85) | (28.09) | (0.77) | | (0.64) | |
| $Ni_2Se_3$ | 31.06 | 25.02 | 0.73 | 0.43 | 0.65 | -0.22 |
| | (30.54) | (28.71) | (0.72) | | (0.68) | |
| $Pd_2Se_3$ | 34.29 | 32.10 | 0.60 | 0.52 | 0.54 | -0.20 |
| | (36.47) | (33.44) | (0.59) | | (0.54) | |

Remarkably, we found that the $M_2Se_3$ monolayers exhibit large out-of-plane NPR in the presence of the strain along the *y* direction ($v_y$), as shown in **Table 3**. In **Figure 6e**, it can be clearly seen that the tensile (compressive) strain along the *y* direction leads to a highly unusual expansion (contraction) in the out-of-plane *z* direction, indicating that three 2D monolayers studied here have the out-of-plane NPR. For 2D materials, the NPR phenomenon was firstly found to exist in the black phosphorus (*b*-P) in 2014[52] and had been confirmed in experiment,[53] which can occur on in-plane[54-58] and out-of-plane[52, 59]. The out-of-plane NPR values in $M_2Se_3$ monolayers are about 4 times larger than that of the borophane (-0.053)[59] and 8 times larger than that of *b*-P (-0.027).[52]

The NPR phenomenon of $M_2Se_3$ monolayers may be attributed to the unique buckling hinge-like structure of $M_2Se_3$ (**Figure 6f** and blue rectangle in **Figure 1a**). When applying the tensile strain along the *y* direction, due to the large positive Poisson's ratio $v_{yx}$, metal atoms *M*1 and *M*4 (*M*2 and *M*3) move inward along the *x* direction, indicated by red arrows in **Figure 6f**. This inward movement reduces the bond angle γ (angle 154 and 263) as shown in **Figure 6d**, which raises the Se-dimer (Se5 and Se6) in the *z* direction. In addition, there is another equivalent hinge-like structure (see **Figure 1c**) which orientates the opposite direction and it lowers its Se-dimer in the -*z* direction with the tensile strain along the *y* direction. Therefore, these opposite movements of Se-dimers, which defines the buckling height, induce unusually large out-of-plane NPR in $M_2Se_3$ monolayer. We can conclude that other metals for 2D $M_2Se_3$ family with the similar buckling hinge-like structure should also have the large out-of-plane NPR values.

In summary, we systemically investigated the structural stabilities, electronic, magnetic and mechanical properties of 2D $M_2Se_3$ monolayers from first-principles method. Our results show that the $Co_2Se_3$ and $Ni_2Se_3$ monolayers have the comparable stability with the recently discovered 2D $Pd_2Se_3$ monolayer. Importantly, $Co_2Se_3$ monolayer exhibits the half-metallity and high-temperature ferromagnetism (>700 K), as confirmed by the MFT and spin-polarized AIMD simulations. Moreover, $M_2Se_3$ monolayers possess the superior mechanical flexibility and large NPR, in which the critical strains and out-of-plane NPR values are much larger than the well-known 2D materials. These outstanding mechanical properties result from the unique buckling structure of $M_2Se_3$ monolayers. The "unfolding" of buckling structure during tensile strain endows the 2D $M_2Se_3$ monolayers with high convertibility and great flexibility. The double agile hinge-like structures and their opposite movements with tensile strain along the $y$ direction induce the giant NPR of $M_2Se_3$ monolayers. It is expected that other similar 2D $M_2Se_3$ monolayers could also have the excellent mechanical properties. High Curie temperature for the ferromagnetism and outstanding flexibility of $Co_2Se_3$ suggest that 2D $M_2Se_3$ family will be promising in the applications of the flexible spintronic nanodevices.

## METHODS

The generalized gradient approximation (GGA) method of Perdewe–Burkee–Ernzerh (PBE)[60] were performed based on density functional theory (DFT) implemented in the Vienna *ab initio* simulation package (VASP).[61, 62] The cutoff energy for the plane wave basis set was taken as 650 eV. Both lattice constants and atomic positions were relaxed until the forces on atoms were less than $10^{-3}$ eV/Å, and the total energy change was less than $10^{-7}$ eV. A vacuum distance of ~18 Å was adopted to avoid the interactions between periodic images. The Brillouin zone integrations were performed by using a Γ-centered grid of 9 × 8 × 1, and the calculations for the density of states (DOS) were performed by using a more accurate grid of

$16 \times 15 \times 1$. The phonon dispersions were calculated with the finite displacement method by using the PHONOPY code[63] with a supercell size of $2 \times 2 \times 1$. The AIMD simulations for the supercell ($3 \times 3 \times 1$) were based on the NVT ensemble with a time step of 1 fs and total time of 5 ps.

The cohesive energy ($E_c$) for 2D $M_2Se_3$ monolayers is defined by:

$$E_c = [E(M_2Se_3) - 4E(M) - 6E(Se)]/10 \tag{2}$$

where $E(M_2Se_3)$, $E(M)$, and $E(Se)$ are the total energy of the $M_2Se_3$ ($M$ = Co, Ni and Pd) monolayer with 10 atoms, the single metal, and single Se atom, respectively. With this definition, a negative $E_c$ indicates an exothermic cohesive.

The in-plane Young's modulus and Poisson's ratio can also be calculated by the following equations:[44, 64]

$$\begin{aligned} E_x &= (C_{11}C_{22} - C_{12}C_{21})/C_{22}, \\ E_y &= (C_{11}C_{22} - C_{12}C_{21})/C_{11}, \\ \upsilon_x &= C_{21}/C_{22}, \\ \upsilon_x &= C_{12}/C_{11}, \end{aligned} \tag{3}$$

## AUTHOR INFORMATION

**Corresponding Author**

*E-mail: hongjw@bit.edu.cn

**Notes**

The authors declare no competing financial interest.

## ACKNOWLEDGEMENTS

This work is supported by the National Science Foundation of China (Grant No. 11572040), the Thousand Young Talents Program of China, and Technological Innovation Project of Beijing Institute of Technology. Theoretical calculations were performed using resources of the National Supercomputer Centre in Guangzhou, which is supported by Special Program for Applied Research on Super Computation of the NSFC-Guangdong Joint Fund (the second phase) under Grant No. U1501501. X.W. acknowledges the National Natural Science Foundation of China (Grant No. 11604011) and Beijing Institute of Technology Research Fund Program for Young Scholars. We thank Prof. J. Jiang for the fruitful discussions.

## ASSOCIATED CONTENT

**Supporting Information.** The Supporting Information available online provides additional tables and figures.

## REFEERENCES

# Supporting Information

**Table S1.** The total energy (meV/f.u.) of NM and AFM states relative to the ferromagnetic FM state of 2D $Co_2Se_3$ monolayer. (AFM2 of $Co_2Se_3$ and all FM and AFM states of $Ni_2Se_3$, $Pd_2Se_3$ are relaxed to NM).

| Structures | FM | NM | AFM1 | AFM2 | AFM3 |
|---|---|---|---|---|---|
| $Co_2Se_3$ | 0 | 100.0 | 104.0 | NM (100.0) | 99.8 |
| $Ni_2Se_3$ | NM | 0 | NM | NM | NM |
| $Pd_2Se_3$ | NM | 0 | NM | NM | NM |

**Table S2.** The elastic constants $C_{ij}$ (GPa) of 2D $M_2Se_3$ monolayer.

| Structures | $C_{11}$ | $C_{12}$ | $C_{22}$ | $C_{66}$ |
|---|---|---|---|---|
| $Co_2Se_3$ | 66.64 | 42.58 | 55.30 | 36.26 |
| $Ni_2Se_3$ | 60.35 | 41.13 | 56.74 | 32.89 |
| $Pd_2Se_3$ | 53.65 | 29.07 | 49.19 | 28.18 |

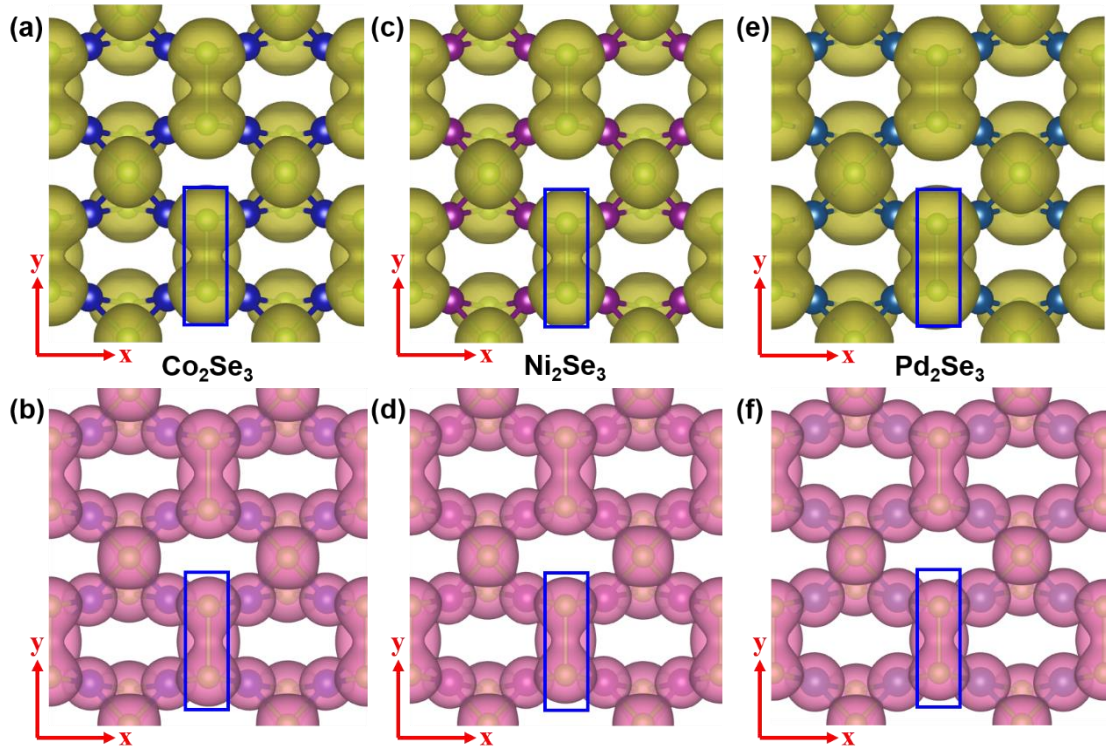

**Figure S1.** Isosurfaces of ELF (a, c, and e) with the value of 0.36 e/Å$^3$ and charge density distribution (b, d, and f) with the value of 0.06 e/Å$^3$ for Co$_2$Se$_3$, Ni$_2$Se$_3$, and Pd$_2$Se$_3$ monolayers. The blue frames indicate the Se dimer.

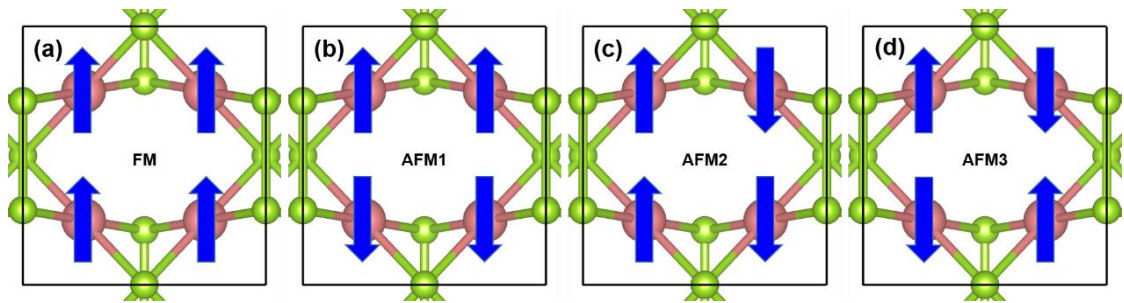

**Figure S2.** The possible magnetic configurations of monolayer $M_2$Se$_3$.

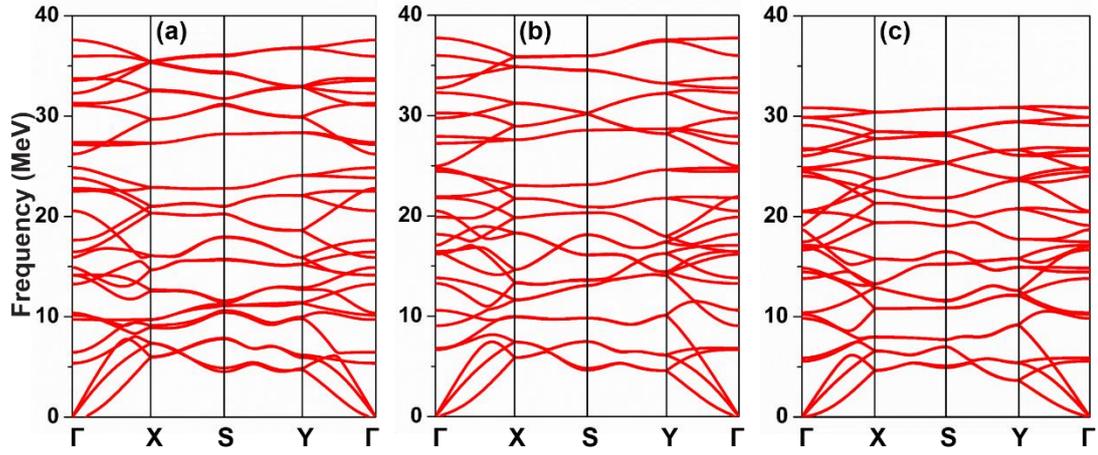

**Figure S3.** The phonon dispersions of the 2D (a) $Co_2Se_3$, (b) $Ni_2Se_3$, and $Pd_2Se_3$ monolayers.

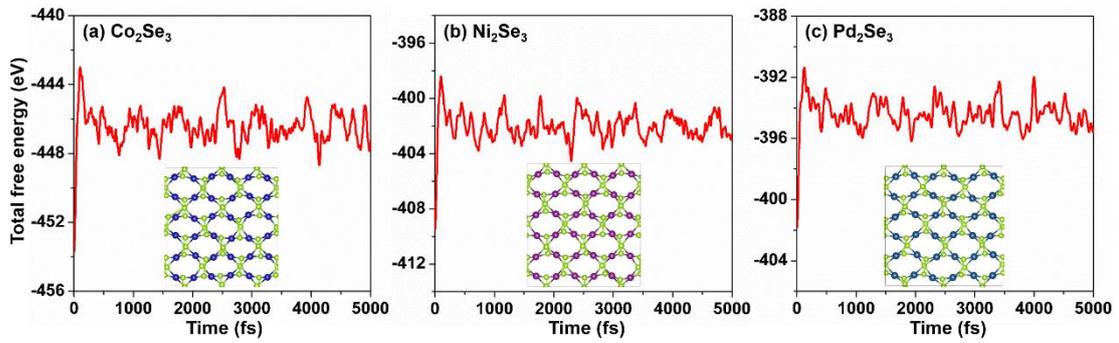

**Figure S4.** The total free energy (eV) fluctuations of the (a) $Co_2Se_3$, (b) $Ni_2Se_3$, and (c) $Pd_2Se_3$ during the AIMD simulations at 600K.

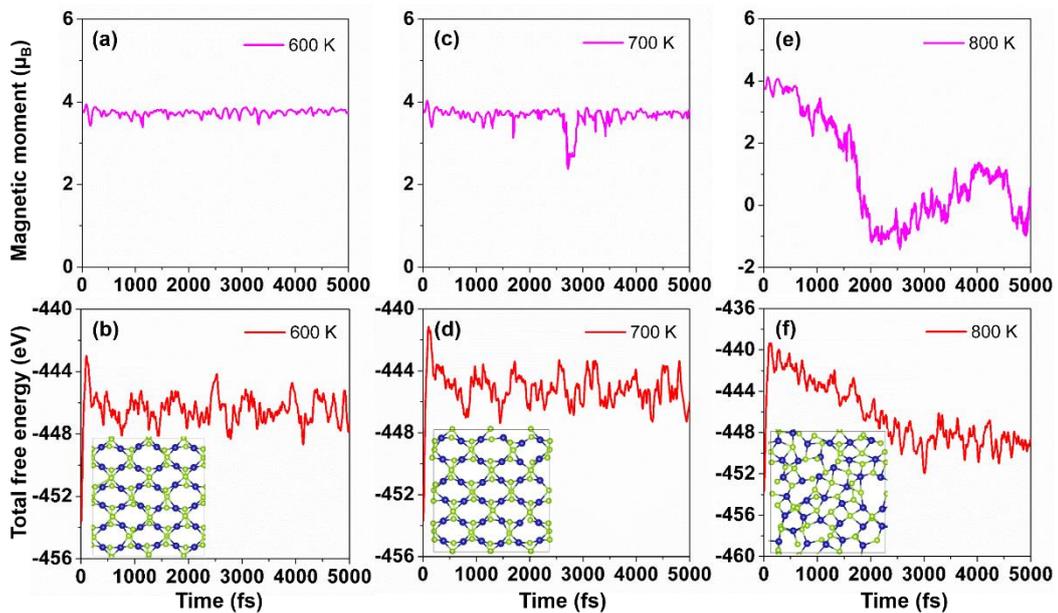

**Figure S5.** The magnetic moment (in $\mu_B$ per unit cell) and total free energy (eV) fluctuations of the $Co_2Se_3$ monolayer under the spin-polarized AIMD simulations at (a, b) 600 K, (c, d) 700 K, and (e, f) 800 K.

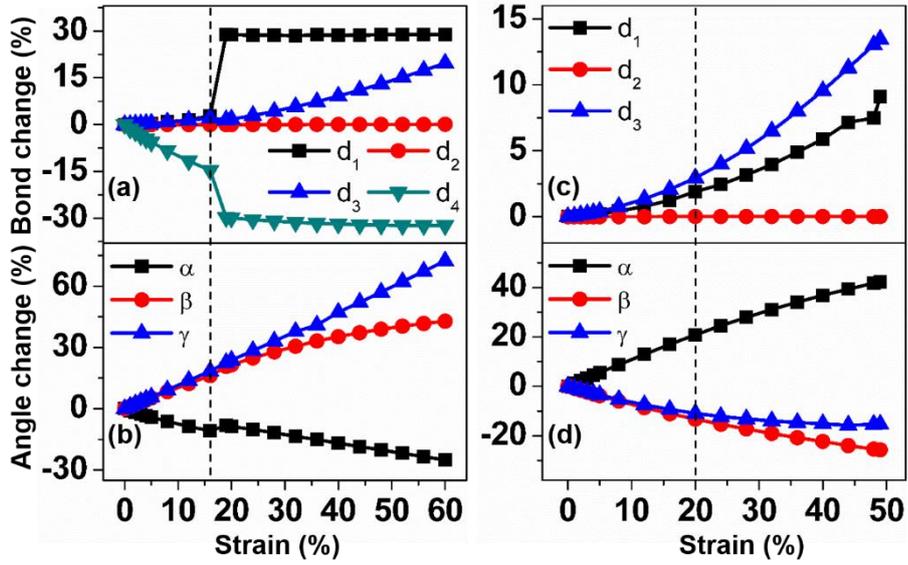

**Figure S6.** The change of bond lengths (atomic distances) and bond angles for the $Ni_2Se_3$ monolayer under tensile strain along the (a, b) $x$ and (c, d) $y$ directions. Here, for the $y$ direction, the $d_4$ is not shown, due to no formation of bond.

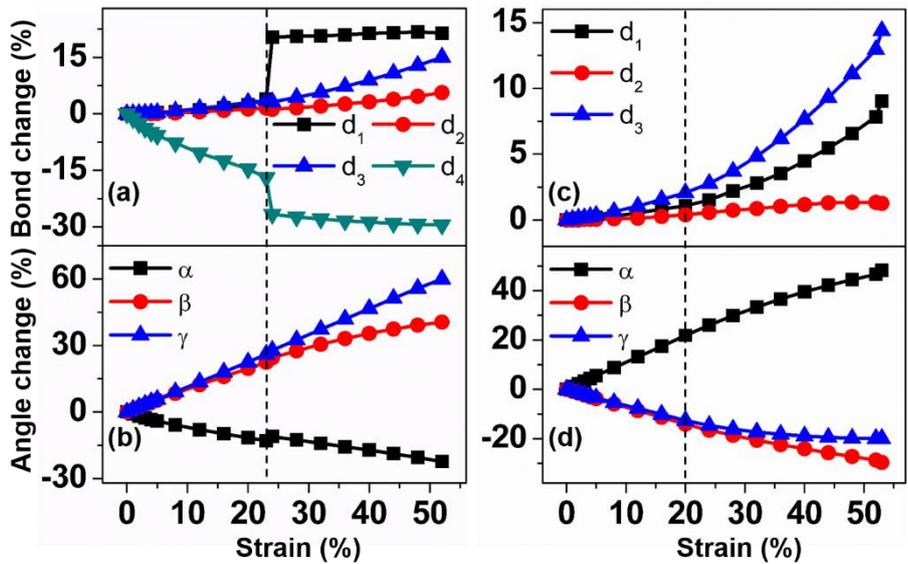

**Figure S7.** The change of bond lengths (atomic distances) and bond angles for the $Pd_2Se_3$

monolayer under tensile strain along the (a, b) *x* and (c, d) *y* directions.

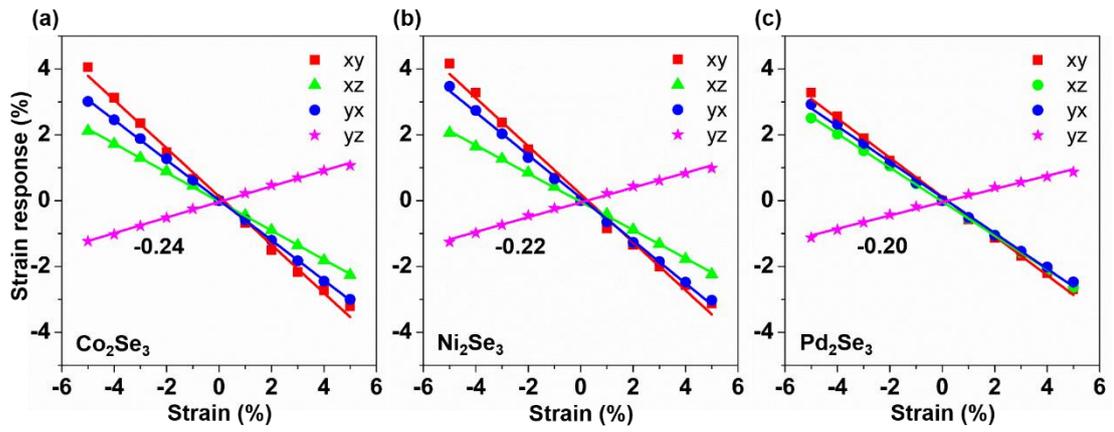

**Figure S8.** The strain response (%) induced by the uniaxial strain for monolayer $M_2Se_3$.